\documentclass[trackchanges]{aastex7}

\usepackage{amsmath}
\usepackage[utf8]{inputenc}  
\usepackage{booktabs}

\begin{document}

\title{MVPinn: Integrating Milne-Eddington Inversion with Physics-Informed Neural Networks for
GST/NIRIS Observations}

\author[orcid=0000-0002-3669-1830,sname='Li']{Qin Li}
\affiliation{Institute for Space Weather Sciences, New Jersey Institute of Technology, 323 Martin Luther King Blvd., Newark, NJ 07102-1982, USA}
\email[show]{ql47@njit.edu}  

\author[orcid=0000-0002-2643-3600,sname='Shen']{Bo Shen} 
\affiliation{Institute for Space Weather Sciences, New Jersey Institute of Technology, 323 Martin Luther King Blvd., Newark, NJ 07102-1982, USA}
\email{bs644@njit.edu}

\author[orcid=0000-0001-6460-408X,sname='Jiang']{Haodi Jiang} 
\affiliation{Computer Science Department, Sam Houston State University, 1905 University Avenue, Huntsville, TX 77340, USA}
\email{bs644@njit.edu}

\author[orcid=0000-0001-9982-2175,sname='Yurchyshyn']{Vasyl B. Yurchyshyn}
\affiliation{Big Bear Solar Observatory, New Jersey Institute of Technology, 40386 North Shore Lane, Big Bear City, CA 92314-9672, USA}
\email{vasyl.yurchyshyn@njit.edu}

\author[sname='Baildon']{Taylor Baildon}
\affiliation{Big Bear Solar Observatory, New Jersey Institute of Technology, 40386 North Shore Lane, Big Bear City, CA 92314-9672, USA}
\email{tab57@njit.edu}

\author[sname='Yi']{Kangwoo Yi}
\affiliation{Institute for Space Weather Sciences, New Jersey Institute of Technology, 323 Martin Luther King Blvd., Newark, NJ 07102-1982, USA}
\email{kangwoo.yi@njit.edu}

\author[orcid=0000-0003-2427-6047,sname='Cao']{Wenda Cao}
\affiliation{Big Bear Solar Observatory, New Jersey Institute of Technology, 40386 North Shore Lane, Big Bear City, CA 92314-9672, USA}
\email{wenda.cao@njit.edu}

\author[orcid=0000-0002-5233-565X,sname='Wang']{Haimin Wang}
\affiliation{Institute for Space Weather Sciences, New Jersey Institute of Technology, 323 Martin Luther King Blvd., Newark, NJ 07102-1982, USA}
\affiliation{Big Bear Solar Observatory, New Jersey Institute of Technology, 40386 North Shore Lane, Big Bear City, CA 92314-9672, USA}
\email{wangha@njit.edu}



\begin{abstract}
We introduce MVPinn, a Physics-Informed Neural Network (PINN) approach tailored for solving the Milne–Eddington (ME) inversion problem, specifically applied to spectropolarimetric observations from the Big Bear Solar Observatory's Near-InfraRed Imaging Spectropolarimeter (BBSO/NIRIS) at the Fe I 1.56 µm lines. Traditional ME inversion methods, though widely used, are computationally intensive, sensitive to noise, and often struggle to accurately capture complex profile asymmetries resulting from gradients in magnetic field strength, orientation, and line-of-sight velocities. By embedding the ME radiative transfer equations directly into the neural network training as physics-informed constraints, our MVPinn method robustly and efficiently retrieves magnetic field parameters, significantly outperforming traditional inversion methods in accuracy, noise resilience, and the ability to handle asymmetric and weak polarization signals. After training, MVPinn infers one magnetogram in about 15 seconds, compared to tens of minutes required by traditional ME inversion on high-resolution spectropolarimetric data.  Quantitative comparisons demonstrate excellent agreement with well-established magnetic field measurements from the SDO/HMI and Hinode/SOT-SP instruments, with correlation coefficients of approximately 90\%. In particular, MVPINN aligns better with Hinode/SOT-SP data, indicating some saturation of HMI data at high magnetic strengths.    We further analyze the physical significance of profile asymmetries and the limitations inherent in the ME model assumption. Our results illustrate the potential of physics-informed machine learning methods in high-spatial-temporal solar observations, preparing for more sophisticated, real-time magnetic field analysis essential for current and next-generation solar telescopes and space weather monitoring.

\end{abstract}

\keywords{\uat{Solar physics}{1476} --- \uat{Magnetogram}{2359} --- \uat{Solar photosphere}{1518} --- \uat{Spectropolarimetry}{1973} --- \uat{Radiative transfer equation}{1336} --- \uat{Solar observatories}{1513} --- \uat{High resolution spectroscopy}{2096}}


\section{Introduction} 

The Milne-Eddington (ME) approximation offers a simplified analytic solution to the radiative transfer equation for polarized light in magnetized stellar atmospheres, represented by the  Unno-Rachkovsky equations \citep{unno, rachkovsky}. Under the assumptions of a plane-parallel geometry with depth-independent physical properties—such as magnetic field strength, velocity, and opacity, the ME approximation reduces the polarized radiative transfer equations to a closed-form solution for the Stokes I/Q/U/V profiles \citep{1998ApJ...506..805B}. Key assumptions include a linear source function that varies with optical depth, a constant line-to-continuum opacity ratio, and spatially uniform magnetic and velocity fields. These simplifications substantially reduce computational complexity by eliminating the need for numerical integration through atmospheres with vertical stratification. Consequently, this makes ME inversions a cornerstone of solar spectropolarimetric analysis, particularly for estimating magnetic field parameters (strength, inclination, azimuth), line-of-sight velocities, and thermodynamic properties \citep{2022A&A...659A.156D}.

The ME application to solar spectropolarimetry gained prominence in the 1980s with the advent of Stokes polarimeters, which required fast inversion methods to interpret magnetic field observations. With development and further tuning, ME inversions then serve as a standard tool for deriving photospheric magnetic fields from lines like Fe I 630.2 nm \citep{Landolfi1982, Skumanich1987}. Since then, various ME-based codes (e.g., MERLIN, VFISV) were integrated into pipelines for major instruments, including the Solar Dynamics Observatory’s Helioseismic and Magnetic Imager \citep[SDO/HMI,][]{SDO_overview, Hmi_VFISV}, and the Hinode Space Telescope’s Spectro-Polarimeter \citep[Hinode/SOT-SP,][]{Hinode_overview, Hinode_VFISV}. While optimized for photospheric lines with weak gradients, ME inversions have also been adapted to chromospheric diagnostics (e.g., He I 1083 nm) under certain slab approximations, which assume a homogeneous layer along the line of sight for simplified radiative transfer modeling \citep{2004A&A...414.1109L}. Despite their limitations in handling complex atmospheres, ME methods remain widely used for their simplicity and practical feasibility for large-scale spectropolarimetric datasets.

There are well-known limitations of the ME model in capturing complex atmospheric dynamics. Depth-dependent gradients in velocity or magnetic fields, with asymmetric or multi-lobed Stokes profiles, cannot be modeled under the ME constant-property assumption \citep{2011LRSP....8....4B}. As a result, ME inversions are more reliable for spectral lines formed in narrow atmospheric layers with weak vertical gradients, such as photospheric Fe I lines (e.g., the 630.2 nm doublet) \citep{2015A&A...577A.140A}. In regions with strong vertical gradients, multi-component atmospheres, or non-local thermodynamic equilibrium (non-LTE), advanced inversion codes like SIR \citep{1992ApJ...398..375R}, NICOLE \citep{2015ascl.soft08002S}, and STiC \citep{2019A&A...623A..74D} are required to model the full depth-dependent structure. These codes solve the RTE iteratively and account for complex atmospheric stratification, at the expense of significantly higher computational cost.

The advent of high-resolution instruments
 presents unique opportunities and challenges. 
(1) ground-based observations are inherently affected by variable seeing, atmopsheric turbulence, instrumental cross-talk, and fluctuating noise levels, which can degrade data quality and complicate the inversion process. (2) Traditional ME inversion methods rely on forward-fitting strategies that are highly sensitive to initial guesses, fitting bounds, and tolerance settings, making them prone to converging on local minima rather than the global minima, especially under such complex observations. (3) Additionally, weak polarization signals and asymmetric Stokes profiles often lead to failed fits or unphysical parameter retrievals.

These limitations call for a smarter, more adaptive inversion strategy—one that can overcome the sensitivity to local minima, failed fits, and noise-related issues inherent in traditional forward-fitting. Physics-Informed Neural Networks (PINNs) offer a promising solution by embedding the ME radiative transfer solution directly into the network’s learning process. Unlike purely data-driven approaches, PINNs incorporate domain-specific physical laws—typically through regularization terms in the loss function to ensure that predicted solutions adhere to governing equations even in the presence of sparse or noisy data \citep{PINN}. By integrating these physical constraints, PINNs mitigate the dependence on initial guesses and fitting bounds that often lead conventional methods into local minima. This hybrid approach has demonstrated success across various fields, from solving Navier–Stokes equations in fluid dynamics \citep{PINN_app_fuild} to modeling stress–strain relationships in materials science \citep{PINN_app_strain}.
In solar physics, PINNs have been applied to infer coronal magnetic fields \citep{2023Jarolim} and have recently been extended to ME inversions using different spectral lines and instruments \citep{2025ApJ...985L...7J}, focusing on single visible-wavelength Fe I 630.2 nm observations from Hinode/SOT-SP and synthetic datasets. Our approach extended PINNs to near-infrared high-spatial-temporal observations, which enables robust, physics-guided magnetic field retrievals that accelerate computation and offer good resilience to the variable conditions inherent in ground-based observations.

 We address the limitations of ME inversions by integrating PINNs directly into the inversion process, so-called Milne-Eddington Inversion with PINN (MVPinn). Our method combines the analytical solution of traditional ME inversions with the flexibility of modern machine learning techniques. By embedding the ME radiative transfer solution directly into the learning process, MVPinn enhances fitting robustness, mitigates sensitivity to initial guesses, and reduces failures due to noise and observational uncertainties (e.g., low-level cross-talk). Specifically tailored for near-infrared spectropolarimetric observations from BBSO/NIRIS, the MVPinn model is trained within approximately two hours, after which it enables rapid inversions at about 15 seconds per 720 $\times$ 720 image. This exhibits a substantial improvement in computational speed, noise tolerance, and physical consistency, making it highly suitable for real-time and high-spatial-temporal solar magnetic field monitoring. 

This paper is structured as follows: Section 2 details the methodology, including the design and training strategy of MVPinn; Section 3 presents the results and validation of our approach; Section 4 discusses the implications and potential limitations; and Section 5 concludes the study and outlines future work.

\section{Methods} \label{sec:style}
\subsection{Fe I 1.56 $\mu$m in Solar Spectropolarimetry}
The Fe I 1.56 $\mu$m lines are a prominent near-infrared doublet used for probing solar magnetic fields. These lines arise from transitions in neutral iron with a high lower-level excitation potential ($\sim$5.4 eV) and have a large effective Land\'{e} g-factors ($\simeq$3.0 for the 15648.5 $\AA$ line and $\simeq$1.5 for the 15652.9 $\AA$ line). The 15648 $\AA$ line in particular is a normal Zeeman triplet with $g_{eff}$=3, one of the highest Land\'{e} factors of any unblended photospheric line from visible through near-IR wavelengths \citep{2016A&A...596A...6L}. Fe I 1565 nm doublet is the most Zeeman sensitive probe of the magnetic field in the deepest photosphere. Several physical factors make the 1.56 $\mu$m lines powerful for magnetic sensing. First, Zeeman splitting $\Delta$$\lambda$ scales roughly as $\Delta\lambda \propto \lambda B$ for a given Land\'{e} factor, so the long wavelength of 1565 nm yields about 2.5 times larger splitting than a visible line around 630 nm, for the same field strength. This amplifies the Stokes V (circular polarization) and Stokes Q/U (linear polarization) signals, thus also improving the detectability of weak fields. Secondly, the H$^{-}$ opacity reaches a minimum in the near-IR around 1.6 $\mu$m, allowing the continuum to form at a deeper formation height—approximately $\sim-50$ km below the $\tau_{500}=1$ surface \citep{2016A&A...596A...2B}. Finally, it provides a clean Zeeman triplet without hyperfine structure or isotopic blends, simplifying its polarization signature.

\subsection{Data and Observations}
The 1.6~m Goode Solar Telescope at Big Bear Solar Observatory (BBSO) operates the Near-InfraRed Imaging Spectropolarimeter (NIRIS), a dual Fabry-P\'{e}rot tunable filter system designed for the 1565~nm lines \citep{2012ASPC..463..291C}. NIRIS can scan the Fe I 1565~nm doublet in $\sim$60 wavelength steps across a round 85'' field of view, recording full Stokes parameters with $\sim$0.083'' pixel scale and $\sim$72s cadence. This allows high-resolution mapping of vector magnetic fields in the deep photosphere. Coupled with the advanced adaptive optics (AO) system at GST, NIRIS regularly achieves diffraction-limited resolution of $\sim$0.2'' to 0.3'', which is essential for detailed studies of small-scale solar features such as sunspot umbral and penumbral fine structures, granulation-scale magnetic elements. Data from NIRIS are inverted using ME inversion pipeline developed at BBSO \citep{2009ASPC..415..101C,2019ASPC..526..317A}. This pipeline converts spectropolarimetric measurements into high-quality vector magnetograms. 

In addition, the cross-instrument comparison were performed using co-temporal magnetograms from SDO/HMI and Hinode/SOT-SP. HMI provides full-disk vector magnetic field measurements at 6173 $\AA$ with a pixel scale of $\sim$1''/pixel and a cadence of 720 seconds \citep{2012SoPh..275..207S}. Hinode/SOT-SP offers high precision spectropolarimetric observation of the Fe I 6301 $\AA$ line with a spatial resolution of $\sim$0.16'' \citep{2008SoPh..249..167T}. These datasets offer complementary perspectives for validating the physical reliability of MVPinn.

\subsection{ME Physics Module implementation}
The ME atmosphere assumes a linear source function with optical depth, a constant magnetic field with height, and no velocity gradients. Specifically, it follows
\begin{equation}
S(\tau) = B_0 + B_1 \tau,   
\end{equation}
where $B_0$ and $B_1$ represent the continuum source function and its gradient with optical depth $\tau$, respectively. This assumption allows an analytic solution to the polarized RTE.
The Zeeman splitting is calculated as:
\begin{equation}
    \nu_B = \frac{g_{\mathrm{eff}} e_0 \lambda_0^2 B}{4 \pi m_e c^2 \Delta\lambda_D}
\end{equation}
where $g_{\mathrm{eff}}$=3.0 as effective Land\'{e} factor, $\lambda_0$=15648.5 $\AA$, $B$ is the magnetic field strength and $\Delta\lambda_D$ is the Doppler width.

For each wavelength point, the Voigt profile describes the absorption line shape as a combination of Gaussian (thermal) and Lorentzian (damping) contributions: 
\begin{equation}
    H(a, v) + i L(a, v) = \frac{1}{\pi} \int_{-\infty}^{\infty} \frac{e^{-y^2}}{(v - y)^2 + a^2} \, dy
\end{equation}
where $H(a, v)$ is the Voigt function and $L(a, v)$ is the dispersion function, $a$ is the damping parameter and $v$ is the normalized frequency. The absorption matrix elements describe how the different Stokes parameters (I, Q, U, V) are affected by the propagation of polarized light through the magnetized atmosphere:
\begin{equation}
    \begin{aligned}
    \eta_I &= 1 + \frac{\eta_0}{2} \left[ \phi_p \sin^2 \theta + \frac{1}{2} (\phi_b + \phi_r)(1 + \cos^2 \theta) \right]   \\
    \eta_Q &= \frac{\eta_0}{2} \left[ \phi_p - \frac{1}{2} (\phi_b + \phi_r) \right] \sin^2 \theta \cos(2\chi)   \\
    \eta_U &= \frac{\eta_0}{2} \left[ \phi_p - \frac{1}{2} (\phi_b + \phi_r) \right] \sin^2 \theta \sin(2\chi)   \\
    \eta_V &= \frac{\eta_0}{2} (\phi_r - \phi_b) \cos \theta.
\end{aligned}
\end{equation}
where $\phi_{p,b,r}$ are Voigt profiles for linear and circular components, $\psi_{p,b,r}$ are dispersion profiles, $\theta$ is the inclination angle, $\chi$ is the azimuth angle, $\eta_0$ is the line-to-continuum opacity ratio. Similarly, the dispersion matrix elements account for the magneto-optical effects that rotate and modify the polarization state:
\begin{equation}
    \begin{aligned}
\varrho_Q &= \frac{\eta_0}{2} \left[ \psi_p - \frac{1}{2} (\psi_b + \psi_r) \right] \sin^2 \theta \cos(2\chi)   \\
\varrho_U &= \frac{\eta_0}{2} \left[ \psi_p - \frac{1}{2} (\psi_b + \psi_r) \right] \sin^2 \theta \sin(2\chi)   \\
\varrho_V &= \frac{\eta_0}{2} (\psi_r - \psi_b) \cos \theta
\end{aligned}
\end{equation}

The full analytic solution to the RTEs for the emergent Stokes profiles is then:
\begin{equation}
    \begin{aligned}
    \Delta &= \eta_I^2(\eta_I^2 - \eta_Q^2 - \eta_U^2 - \eta_V^2 + \varrho_Q^2 + \varrho_U^2 + \varrho_V^2) - (\eta_Q \varrho_Q + \eta_U \varrho_U + \eta_V \varrho_V)^2   \\
    I &= B_0 + B_1 \frac{ \eta_I (\eta_I^2 + \varrho_Q^2 + \varrho_U^2 + \varrho_V^2) }{ \Delta }   \\
    Q &= -B_1 \frac{\eta_I^2 \eta_Q + \eta_I(\eta_V \varrho_U - \eta_U \varrho_V) + \varrho_Q (\eta_Q \varrho_Q + \eta_U \varrho_U + \eta_V \varrho_V)}{\Delta}   \\
    U &= -B_1 \frac{\eta_I^2 \eta_U + \eta_I(\eta_Q \varrho_V - \eta_V \varrho_Q) + \varrho_U (\eta_Q \varrho_Q + \eta_U \varrho_U + \eta_V \varrho_V)}{\Delta}   \\
    V &= -B_1 \frac{\eta_I^2 \eta_V + \eta_I(\eta_U \varrho_Q - \eta_Q \varrho_U) + \varrho_V (\eta_Q \varrho_Q + \eta_U \varrho_U + \eta_V \varrho_V)}{\Delta}.
\end{aligned}
\end{equation}

\subsection{Training and Loss Design}
MVPinn was tailored for the NIRIS inversion task, involving up to 2 hours of training on Stokes profiles from a single time point, followed by 15-second inference for the remaining data throughout the day. The training and validation dataset consisted of selected calibration files obtained from the full Stokes measurements on 2024-07-25, comprising a total of 230 datasets. Each dataset included spectropolarimetric profiles (Stokes I, Q, U, V) sampled from -3 to +3 $\AA$ around 1.56 $\mu$m, at wavelength points with a step size of 0.15 $\AA$. For this study, a single calibrated Stokes profile at 2024-07-25 16:32:22 UT was selected for training, and 16:51:51 UT was selected for comparative analysis. An animation of the full-day inversion results was also provided. 
The input layer consisted of the Stokes profiles (I, Q, U, V), The hidden layer employed the hyperbolic tangent (tanh) activation function. The tanh function was chosen for its symmetric, bounded nature. The output layer was explicitly configured to predict 9 physical parameters: the magnetic field strength ($B$), inclination angle ($\theta$), and azimuth angle ($\chi$), line-to-continuum opacity ratio ($\eta_0$), doppler width ($\Delta\lambda_D$), damping parameter ($a$), and line center shift ($\Delta\lambda_0$). Physical bounds were incorporated directly into this output layer, ensuring the outputs fall within the bounds strictly: 


\begin{equation}
\left[
\begin{array}{c}
B \\
\theta \\
\chi \\
\eta_0 \\
\Delta\lambda_D \\
a \\
\Delta\lambda_0
\end{array}
\right]
\in
\begin{array}{c}
{[0,\ 5000]} \\
{[0,\ \pi]} \\
{[0,\ \pi]} \\
{[0.5,\ 20]} \\
{[0.12,\ 0.25]} \\
{[0,\ 10]} \\
{[-0.25,\ 0.25]}
\end{array}.
\end{equation}

\begin{figure}[ht!]
\centering
\vspace{-0mm}
\includegraphics[width=1\linewidth]{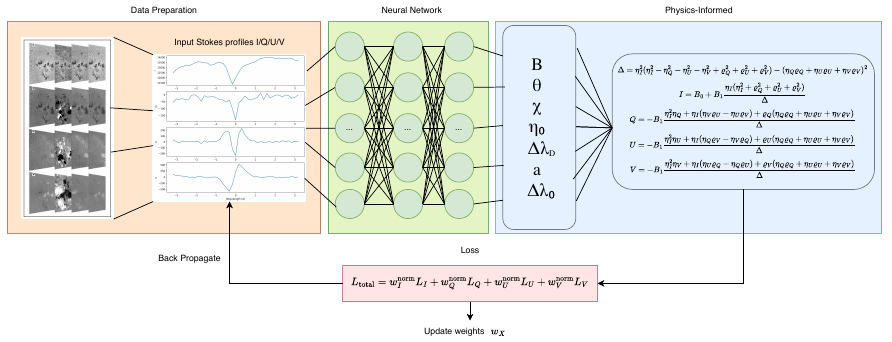}
\caption{Training workflow of MVPinn. The input Stokes profiles (I, Q, U, V) are processed through a physics-informed neural network that directly predicts ME atmospheric parameters, including magnetic field strength ($B$), inclination ($\theta$), azimuth ($\chi$), line-to-continuum opacity ratio ($\eta_0$), Doppler width ($\Delta\lambda_D$), damping parameter ($a$), and line center shift ($\Delta\lambda_0$). These predicted parameters are passed through an analytical ME forward model to synthesize Stokes profiles, which are compared to the observed profiles to compute the loss. The training process iteratively minimizes this loss through backpropagation to update network weights.}
\label{fig:PINN}
\end{figure}

The MVPinn was trained over a total of 60 epochs, a number selected to balance convergence quality with computational efficiency. A batch size of 256 pixels was used throughout the training, where each batch was randomly sampled from the available 720 $\times$ 720 pixel images to spatially vary profiles and to avoid overfitting to localized features. This random sampling process ensures that the network experiences a wide variety of stokes profiles across the dataset in each training iteration. The training workflow is shown in Figure \ref{fig:PINN}. During training, the network predicts the ME physical parameters directly from the input Stokes profiles (I, Q, U, V). These inferred parameters are passed through an analytical ME forward model, which synthesizes corresponding Stokes profiles based on the predicted atmospheric conditions. The loss function minimizes the difference between these synthesized profiles and the observed Stokes profiles, effectively embedding the ME radiative transfer solution into the learning process.

To ensure generalizability and prevent overfitting to specific regions within the training data, we applied L2 regularization with a weight decay coefficient of 10$^{-5}$. For optimization, we employed the Adam optimizer \citep{kingma2017adammethodstochasticoptimization}, a well-established adaptive algorithm known for its effective convergence. The training was accelerated using GPU computation, which significantly reduced total processing time and made the method scalable for high-spatial-temporal-resolution, large-volume datasets typical of modern solar observations. 

The final loss is formulated as a weighted sum of the residuals between the synthesized and observed Stokes profiles, combining the data loss from all four Stokes components (I, Q, U, V) as follows
\begin{equation}
L_{\mathrm{total}} = w_I^{\mathrm{norm}} L_I + w_Q^{\mathrm{norm}} L_Q + w_U^{\mathrm{norm}} L_U + w_V^{\mathrm{norm}} L_V,
\end{equation}
where $w_X^{\mathrm{norm}}$ represents the normalized adaptive weights and $L_X$ represents the loss between observed and predicted stokes profiles, in which $X$ represents for I, Q, U, V components. To ensure that each Stokes parameter contributes appropriately to the total loss, their individual losses are scaled by normalized weights $w_x$, which account for differences in signal magnitude and noise levels among the Stokes components. This normalization prevents the typically stronger Stokes I signal from dominating the loss function and ensures that the polarization signals (Q, U, V) are properly emphasized during training. By balancing the contributions from each Stokes component, the network is guided toward fitting the full spectropolarimetric profile consistently across all channels.

For each component (I, Q, U, V), the loss is calculated as the mean squared error (MSE) normalized by the variance:
\begin{equation}
    \begin{aligned}
L_I &= \frac{1}{N \sigma_I^2} \sum_{i=1}^{N} \left( I_{\mathrm{pred},i} - I_{\mathrm{obs},i} \right)^2  \\
L_Q &= \frac{1}{N \sigma_Q^2} \sum_{i=1}^{N} \left( Q_{\mathrm{pred},i} - Q_{\mathrm{obs},i} \right)^2  \\
L_U &= \frac{1}{N \sigma_U^2} \sum_{i=1}^{N} \left( U_{\mathrm{pred},i} - U_{\mathrm{obs},i} \right)^2  \\
L_V &= \frac{1}{N \sigma_V^2} \sum_{i=1}^{N} \left( V_{\mathrm{pred},i} - V_{\mathrm{obs},i} \right)^2
\end{aligned}
\end{equation}
where $N$ is the number of wavelength points; in this study, it is 31 after data reduction. $\sigma_X^2$ is the variance of the observed Stokes component $X$, $X_{\mathrm{pred},i}$ and $X_{\mathrm{obs},i}$ are the predicted and observed values at wavelength i.

The adaptive weights are calculated using gradient norms with power scaling:
\begin{equation}
w_X = \frac{1}{\left( \| \nabla L_X \|^{1.5} + \epsilon \right)}
\end{equation}
where $\epsilon$=$10^{-6}$ for numerical stability. 

The weights are then normalized to sum to 1, and finally smoothed with Exponential Moving Average (EMA) following
\begin{equation}
w_X^{\mathrm{norm}} = \frac{w'_X}{\sum_{Y \in \{I, Q, U, V\}} w'_Y}
\end{equation}

\begin{equation}
w_X^{t} = \beta w_X^{t-1} + (1 - \beta) w_X^{\mathrm{norm}}
\end{equation}
where the smoothing factor $\beta$=0.9, $t$ is the current iteration. 

\section{Results}
\subsection{Vector Magnetic Fields: MVPinn versus ME inversion}\label{subsection: overall comparison}

\begin{figure}[ht!]
\centering
\includegraphics[width=1\linewidth]{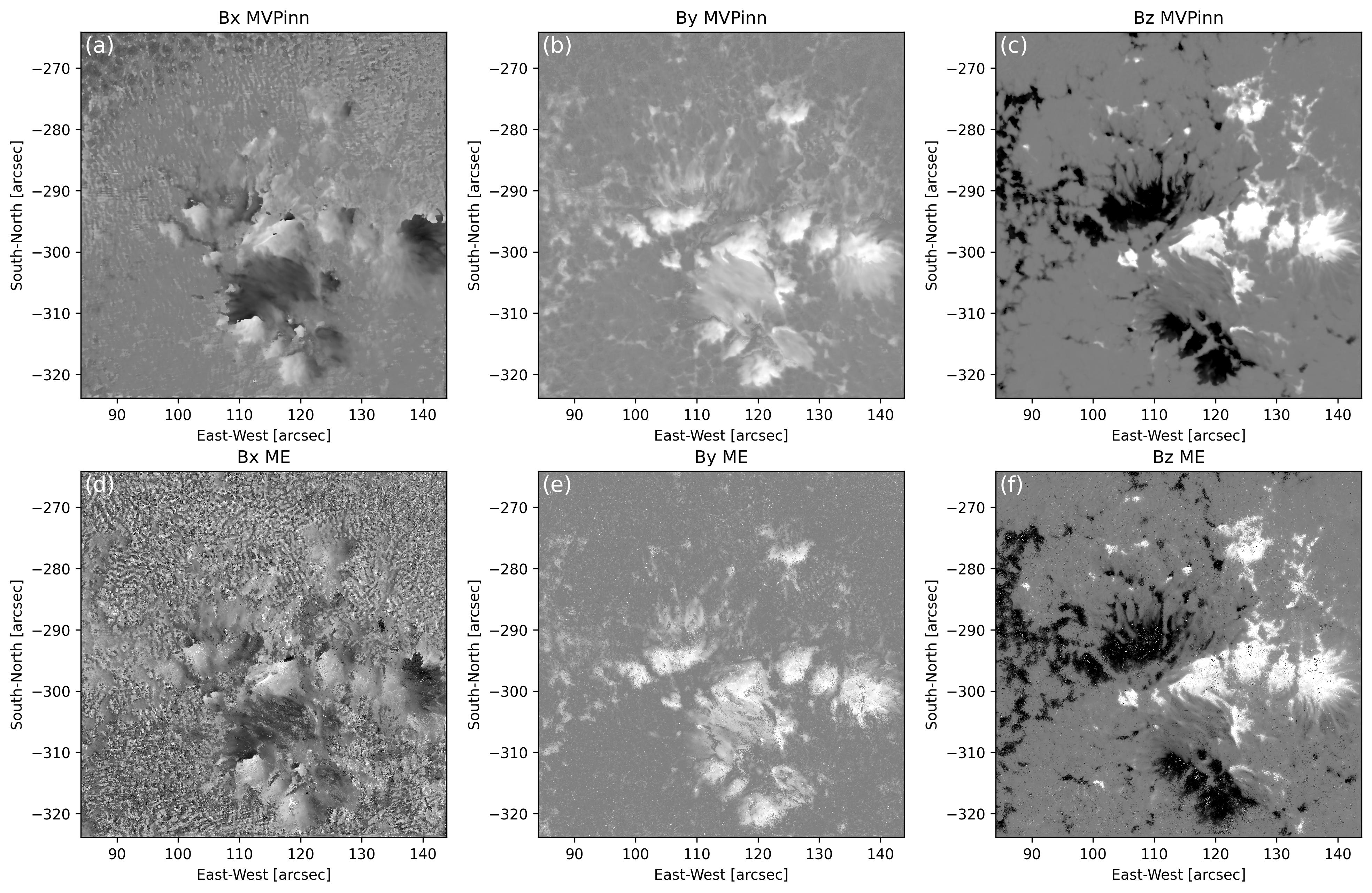}
\caption{Comparison of vector magnetic fields observed on 2024-07-25 at 16:51:51 UT, inferred using our MVPinn method versus the traditional ME inversion. Panels (a-c) show the vector magnetic field components $B_x$, $B_y$, and $B_z$, respectively, obtained from the MVPinn method. Panels (d-f) show the vector magnetic field components $B_x$, $B_y$, and $B_z$, derived using the ME inversion. All magnetograms are presented on a scale ranging from -1500 Gauss to +1500 Gauss. As the 180-degree of transverse field is not resolved,   one of the components (By) is always assigned positive values.}
\label{fig:Bfield}
\end{figure}

\begin{figure}[ht!]
\centering
\includegraphics[width=1\linewidth]{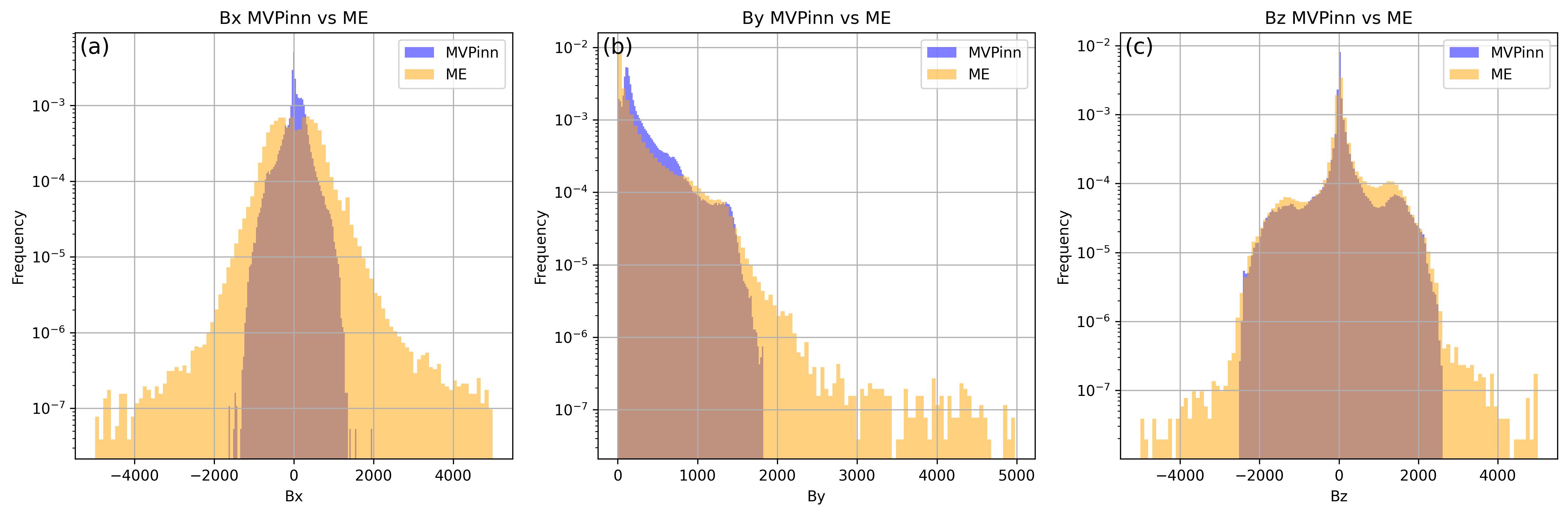}
\caption{Statistical distribution analysis of the vector magnetic field components $B_x$, $B_y$, and $B_z$ inferred from the MVPinn and ME inversion methods. The histograms are plotted on a logarithmic scale and show the normalized distribution of (a) $B_x$, (b) $B_y$, and (c) $B_z$. The MVPinn results are shown in blue, while ME inversion results are shown in orange.}
\label{fig:distribution}
\end{figure}

We first investigated the effectiveness of MVPinn inversion by comparing the retrieved vector magnetic field components ($B_x$, $B_y$, $B_z$) with those obtained from traditional ME inversion. Figure \ref{fig:Bfield} provides a visualization of these fields from an active region observed on 2024-07-25 at 16:51:51 UT. MVPinn consistently reproduced clearer and more structurally coherent magnetic field maps compared to the ME inversion. In particular, the horizontal magnetic field components $B_x$ and $B_y$ derived using MVPinn demonstrated significantly reduced noise levels and smoother transitions across complex magnetic features. 

Quantitatively, the comparison of the histogram distributions of the inferred magnetic fields, shown in Figure \ref{fig:distribution}, reinforces the visual assessment. MVPinn yielded tighter, narrower distributions of magnetic field strengths, especially for $B_x$ and $B_y$, which implies that MVPinn regularize the inversion outcomes and minimize noise-induced outliers. By contrast, the ME inversion displayed broader tails, indicative of greater susceptibility to observational noise and a higher probability of generating unrealistic magnetic field values. A detailed comparison from other instruments are presented in the section 3.3.

\subsection{Detailed Analysis of Stokes Profile Fits}\label{subsection: Line Profile analysis}

\begin{figure}[ht!]
\centering
\includegraphics[width=1\linewidth]{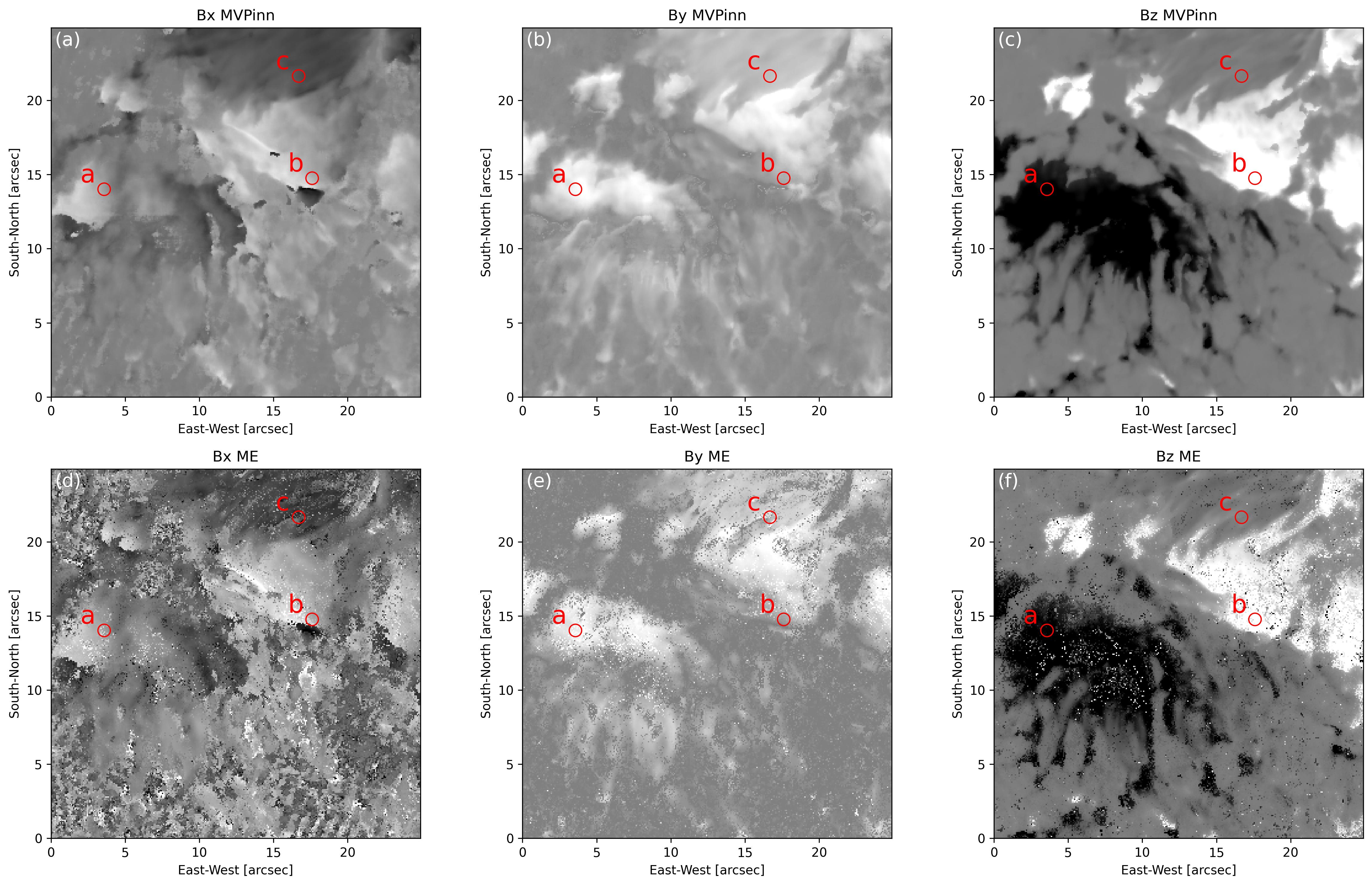}
\caption{Comparison of difference inference from MVPinn and ME inversion on a smaller field of view. Three locations were selected for detailed stokes profile analysis in Section. \ref{subsection: Line Profile analysis} (a) strong magnetic magnitude in negative polarity. (b) strong magnetic magnitude in positive polarity (c) highly dynamic penumbral region. Three locations are selected for comparison of strong vertical or transverse magnetic components, without considering noises.  The tails of ME results are most likely due to inversion errors and noises.}
\label{fig:Bfield_fov_01}
\end{figure}

The performance of MVPinn is directly evaluated based on its fitting accuracy to the observed Stokes profiles. Figure \ref{fig:Bfield_fov_01} presents a comparison between the observed Stokes profiles (I, Q, U, V) and those synthesized from three inversion approaches: MVPinn, traditional ME inversion, and the Weak Field Approximation (WFA). The comparison is carried out for both consistent and divergent cases, as shown in the following figures. Table~\ref{tab:combined_mvpinn_me} summarizes the inferred parameter values at these locations. The consistent cases include three representative points, indicated in Figure~\ref{fig:Bfield_fov_01}: (a) a region of strong magnetic field within positive polarity, (b) a region of strong magnetic field within negative polarity, and (c) a penumbral region characterized by highly transverse magnetic field components and dynamic fine structures.

In the strong-field regions (locations a and b), both MVPinn and traditional ME inversion produce broadly consistent results, particularly where the polarization signals are strong. In Region (a), the difference in $B_x$ is moderate (-112 G), while the differences in $B_y$ and $B_z$ are 153 G and 97 G, respectively. These values are relatively small compared to the total field strength ($>2500 G$), suggesting good agreement between the two methods. Region (b) shows even closer agreement in all three components, with differences of -49 G in $B_x$, 39 G in $B_y$, and -99 G in $B_z$. Generally, MVPinn consistently demonstrates slightly superior performance in fitting asymmetric profiles, particularly in Stokes V, where it more accurately captures subtle asymmetries and fine-scale structure in the observed profiles. In contrast, traditional ME inversions tend to produce fits that deviate slightly from the observed data, especially where asymmetries are present, particularly in Stokes I.

\begin{table}[ht]
\centering
\caption{Comparison of inferred parameters from MVPinn and ME inversion at six representative positions (a–f). Results from MVPinn are shown outside the parentheses, and results from ME inversion are in parentheses. The parameters include magnetic field strength ($B$), inclination angle ($\theta$), azimuth angle ($\chi$), line-to-continuum opacity ratio ($\eta_0$), Doppler width ($\Delta\lambda_D$), damping parameter ($a$), and line center shift ($\Delta\lambda_0$). $B_0$ and $B_1$ represent the source function and its gradient, respectively. $B_x$, $B_y$, and $B_z$ are the vector magnetic field components, derived from the inferred parameters for comparison.}
\begin{tabular}{lcccccc}
\hline
& \multicolumn{3}{c}{\textbf{Consistent Cases}} & \multicolumn{3}{c}{\textbf{Divergent Cases}} \\
\cmidrule(lr{0.2em}){2-4}
\cmidrule(lr{0.2em}){5-7}
Parameter & (a) & (b) & (c) & (d) & (e) & (f) \\
\hline
$B$ (G) & 2702.95 (2732.98) & 2063.85 (2141.23) & 1033.70 (572.71) & 2072.63 (1746.29) & 504.59 (1535.53) & 119.46 (789.70) \\
$\theta$ (rad) & 2.69 (2.75) & 0.58 (0.55) & 1.60 (1.67) & 2.45 (1.26) & 2.22 (3.10) & 1.55 (1.73) \\
$\chi$ (rad) & 1.53 (1.41) & 1.11 (1.06) & 2.24 (1.82) & 0.84 (0.88) & 1.78 (0.00) & 1.61 (0.07) \\
$\eta_0$ & 1.32 (15.05) & 8.92 (16.22) & 5.99 (19.81) & 0.72 (19.99) & 19.45 (0.57) & 19.66 (0.51) \\
$\Delta\lambda_D$ (\AA) & 0.12 (0.12) & 0.12 (0.12) & 0.14 (0.12) & 0.12 (0.25) & 0.25 (0.25) & 0.12 (0.25) \\
$a$ & 1.84 (4.07) & 0.91 (2.92) & 1.02 (5.39) & 1.16 (2.21) & 0.31 (1.89) & 0.07 (3.15) \\
$\Delta\lambda_0$ (\AA) & -0.04 (-0.04) & -0.01 (-0.01) & 0.02 (-0.06) & -0.05 (-0.16) & -0.04 (-0.07) & 0.01 (-0.25) \\
$B_0$ & -0.12 (0.14) & 0.56 (0.31) & 0.44 (0.22) & -0.40 (0.73) & 0.91 (0.56) & 0.81 (-0.60) \\
$B_1$ & 0.83 (0.56) & 0.25 (0.50) & 0.57 (0.75) & 1.26 (0.12) & 0.07 (0.40) & 0.20 (1.60) \\
$B_x$ (G) & 51.80 (164.01) & 504.83 (553.94) & -640.78 (-139.92) & 881.41 (1056.15) & -83.80 (60.11) & -4.20 (778.38) \\
$B_y$ (G) & 1186.97 (1033.61) & 1016.38 (977.16) & 810.43 (552.68) & 977.62 (1284.20) & 392.89 (0.00) & 119.36 (51.95) \\
$B_z$ (G) & -2427.83 (-2524.66) & 1723.83 (1822.96) & -33.80 (-54.48) & -1600.99 (533.76) & -305.33 (-1534.35) & 2.79 (-122.64) \\
\hline
\end{tabular}
\label{tab:combined_mvpinn_me}
\end{table}
\begin{figure}[ht!]
\centering
\includegraphics[width=1\linewidth]{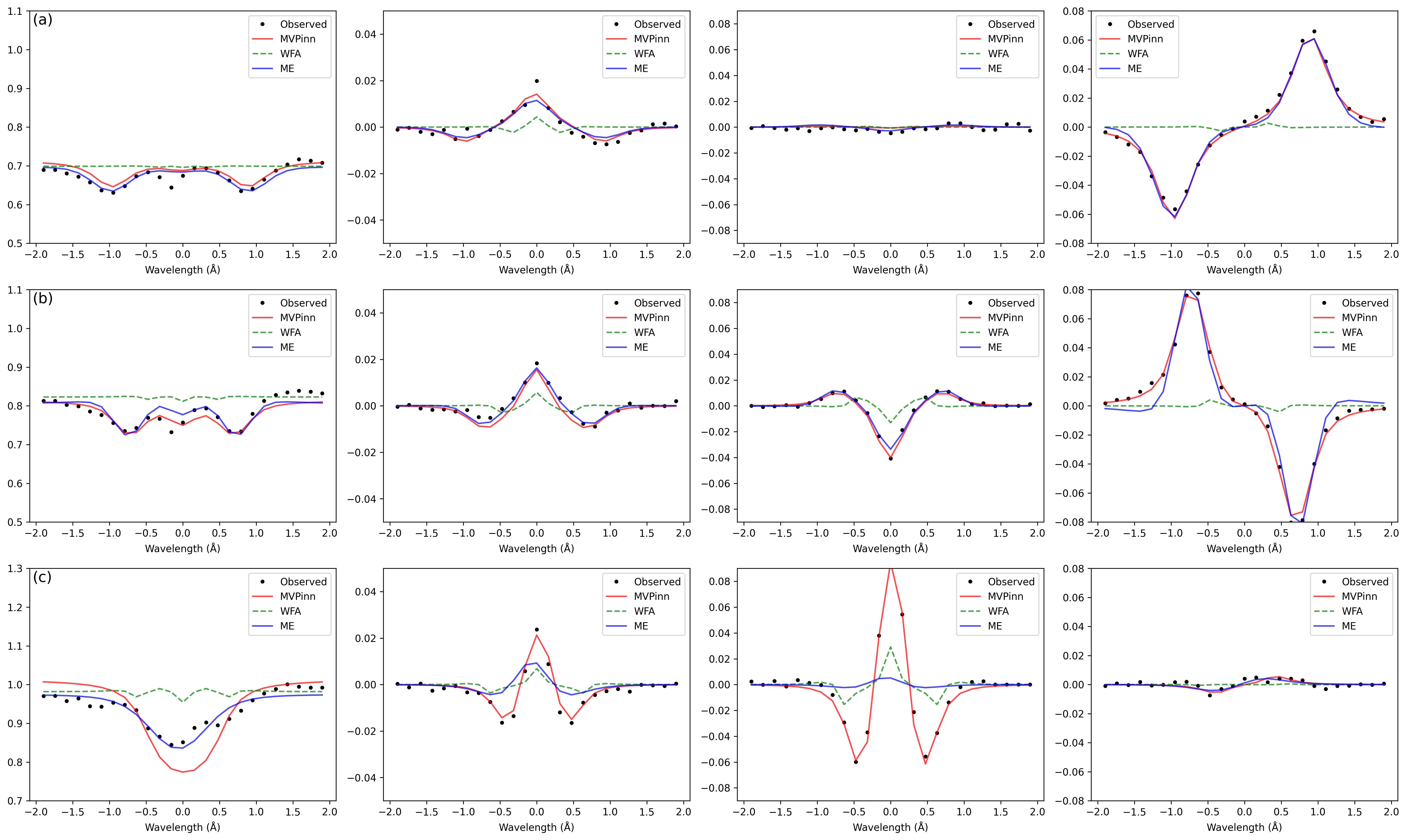}
\caption{Comparison of Stokes profiles and inversion results corresponds to three locations selected from (a,b,c) shown above, inferred from different inversion methods: MVPinn (red), ME (blue), WFA (also used as initial condition for ME, green). Panels from left to right in each row display the normalized Stokes I, Q, U, and V profiles.}
\label{fig:stokes_fitted_01}
\end{figure}
In Region (c), a penumbral location where the transverse components dominate, the discrepancy between the methods becomes more pronounced. The differences reach -501 G in $B_x$ and 258 G in $B_y$, while $B_z$ shows a smaller difference of 21 G. This is consistent with expectations, as the penumbral region is dominated by transverse fields, and traditional ME inversions often struggle to robustly fit Q and U profiles in such regions. MVPinn appears to capture the transverse field with much higher sensitivity and possibly more physical realism. Traditional ME inversion shows reduced robustness in this scenario, often prioritizing the fit to the Stokes I/V profile while failing to adequately reproduce the Q and U profiles. In contrast, MVPinn demonstrates a more balanced and consistent fitting across all Stokes parameters, suggesting a more flexible and physically realistic retrieval of the underlying magnetic configuration. Notably, in all cases, the inferred parameters remain well within their physical bounds, indicating successful convergence and physically plausible solutions across all methods.

\begin{figure}[ht!]
\centering
\includegraphics[width=1\linewidth]{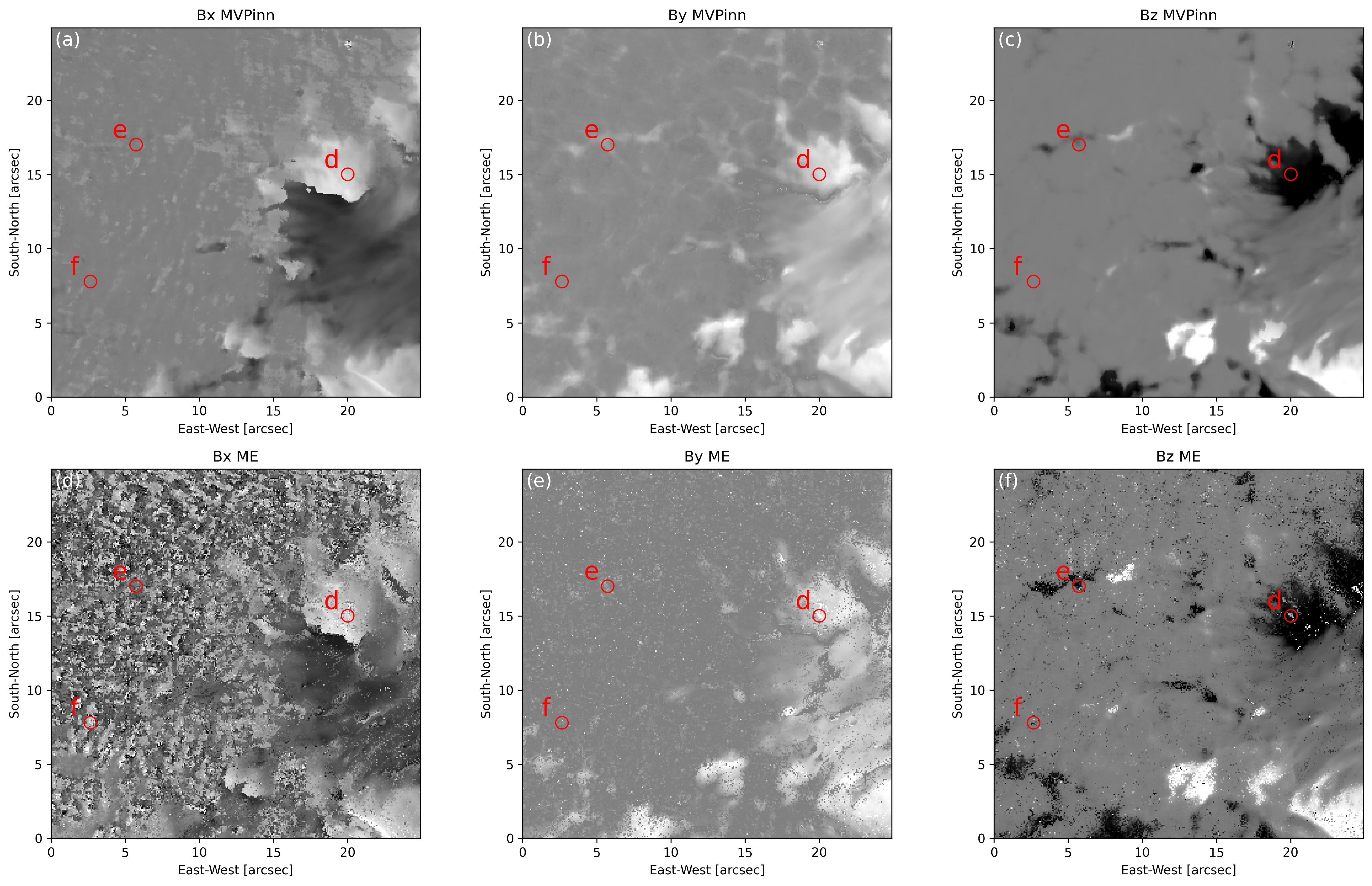}
\caption{Comparison of difference inference from MVPinn and ME inversion on a smaller field of view. Three locations were selected for detailed stokes profile analysis in Section. \ref{subsection: Line Profile analysis} (a) one of noisy point regularly found within sunspot. (b) regrepents discrepency of magnetic field strength at some small and moderate magnetic elements, and (c) one of noisy points regularly found within quiet sun. The purpose is to investigate the physical integrity of these kind of regions about their inference.}
\label{fig:Bfield_fov_02}
\end{figure}

\begin{figure}[ht!]
\centering
\includegraphics[width=1\linewidth]{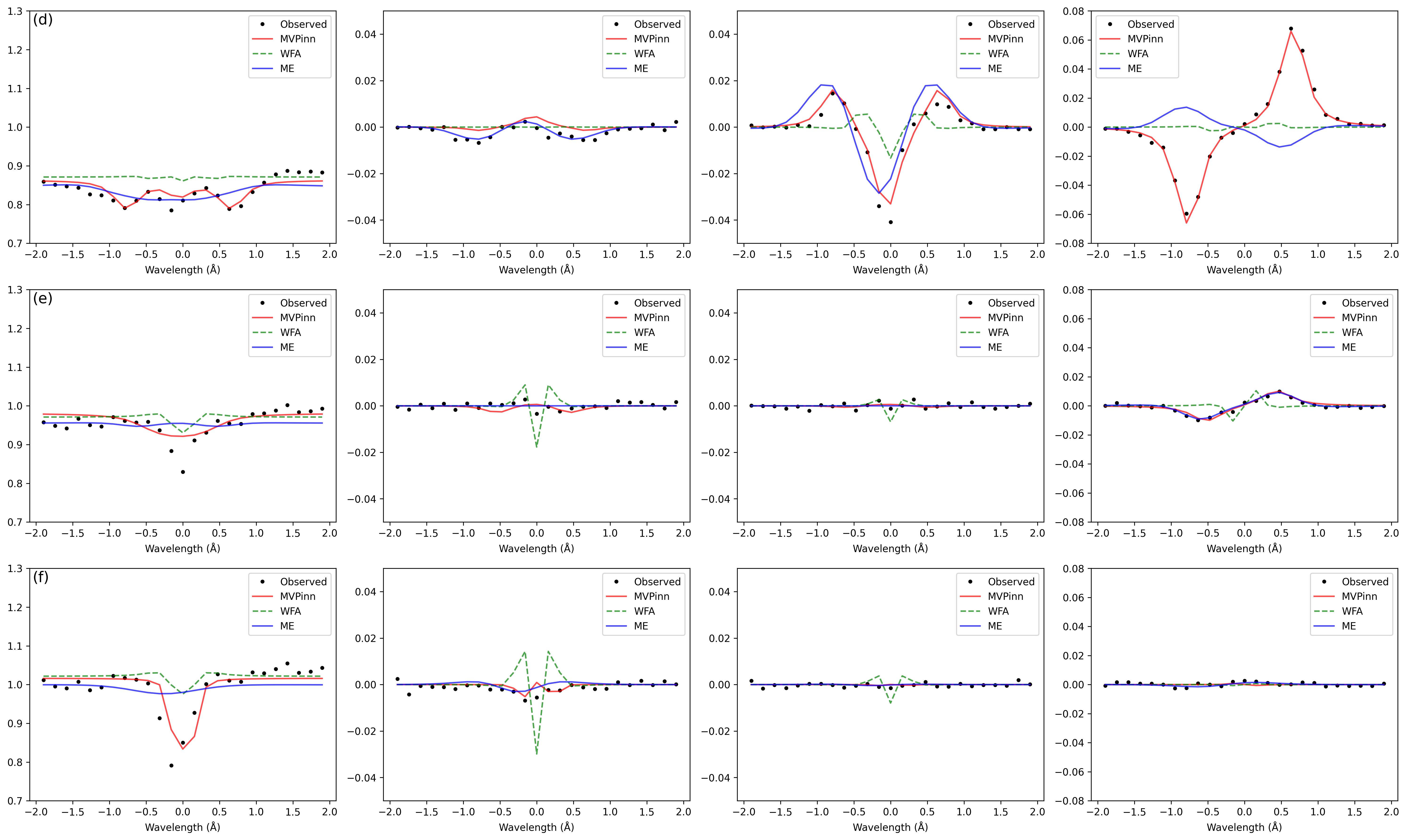}
\caption{Comparison of Stokes profiles and inversion results corresponds to three locations selected from (d,e,f) shown above, inferred from different inversion methods: MVPinn (red), ME (blue), WFA (also used as initial condition for ME, green). Panels from left to right in each row display the normalized Stokes I, Q, U, and V profiles.}
\label{fig:stokes_fitted_02}
\end{figure}

Following the previous analysis focused on positive and negative strong magnetic field regions, we also selected three locations where there are clear discrepancies between the two inversion methods, as shown in Figure \ref{fig:Bfield_fov_02}. These locations are selected to explore ambiguous fitting scenarios: (d) a noisy point within a strong negative magnetic field region, (e) a location showing an obvious discrepancy in field strength within a moderate magnetic feature, and (f) a representative noisy point commonly found in the quiet Sun. The corresponding Stokes profiles and fitting results are shown in Figure \ref{fig:stokes_fitted_02}. The purpose is to investigate the fitting differences and the underlying reasons behind these discrepancies.

In Region (d), a location affected by noisy or ill-constrained signals, ME inversion performs poorly. The retrieved $\eta_0$ from ME hits the upper bound, suggesting the solution terminated at a constraint. Here, the differences between methods are substantial: -175 G in $B_x$, -307 G in $B_y$, and over -2135 G in $B_z$. Visual inspection confirms that MVPinn provides a better fit to the observed Stokes profiles, while the ME solution exhibits large deviations, especially in Stokes V, indicating a failure to converge to a physically meaningful solution.

Region (e) highlights a case of overestimation by the ME method. MVPinn retrieves $B_x$ = -83.80 G, $B_y$ = -392.89 G, and $B_z$ = -305.33 G, whereas ME infers $B_x$ = 60.11 G, $B_y$ = 0.00 G, and $B_z$ = -1534.35 G. The differences in $B_x$ and $B_y$ are -144 G and 393 G, respectively, but the most significant discrepancy is found in $B_z$, where ME overestimates the line-of-sight component by more than 1200 G. This likely stems from the ME method's tendency to deprioritize Stokes I when Q and U signals are weak, whereas MVPinn maintains balanced fitting across all Stokes components. In this case, the discrepancy is not due to noise but highlights different fitting priorities between the two methods.

Finally, in Region (f), representing a weak field location, MVPinn retrieves low field values $B_x$ $\approx$ -4.2G, $B_y$ $\approx$ 119.4 G, $B_z$ $\approx$ 2.8 G. In contrast, ME inversion reports unrealistically high values: $B_x$ $\approx$ 778.4 G, $B_y$ $\approx$ 52.0 G, $B_z$ $\approx$ -122.6 G, resulting in differences of -783 G, 67 G, and 125 G for the respective components.  These differences suggest that ME, especially in low-signal regions, may easily overfit noisy residuals, particularly when it neglects to fit the Stokes I profile. MVPinn, on the other hand, adapts to the noisy observational conditions and produces solutions that are more realistic for weak field regions.

\begin{figure}[ht!]
\centering
\includegraphics[width=1\linewidth]{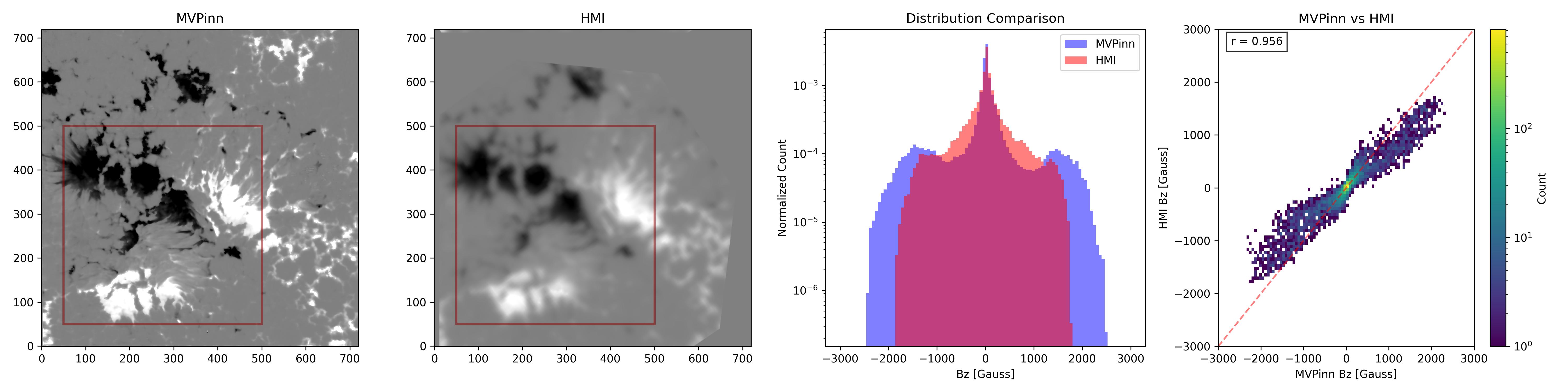} \\
\vspace{-1mm}
\includegraphics[width=1\linewidth]{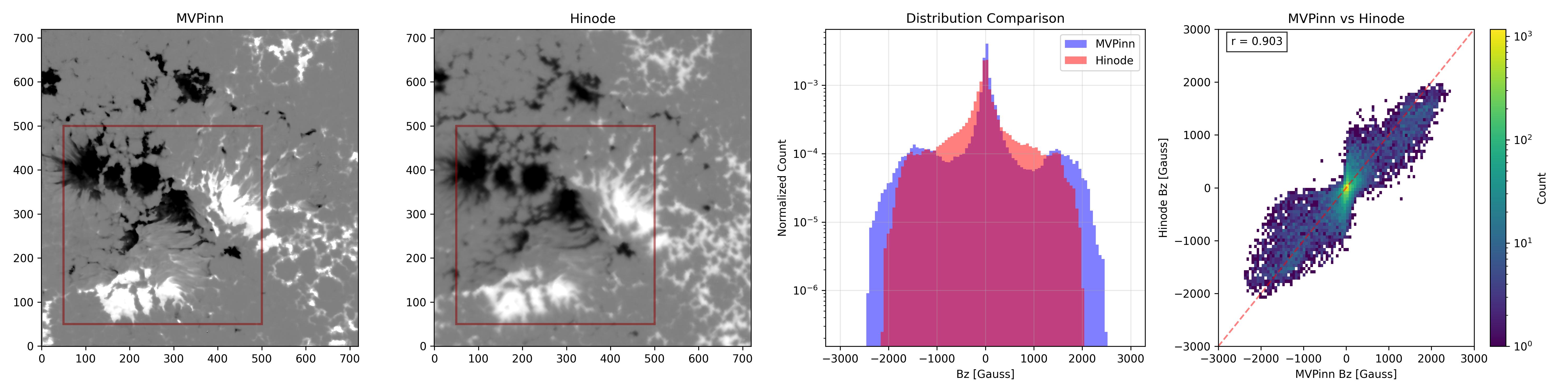}
\caption{A comparative analysis of the vertical magnetic component $B_z$ obtained from MVPinn inversions with aligned co-temporal observations from SDO/HMI and Hinode/SOT-SP, as well as the normalized histograms of $B_z$ from both datasets and pixel-to-pixel comparison of $B_z$ between MVPinn and other instruments, along with Pearson correlation coefficient $\gamma$.}
\label{fig:stacked_images}
\end{figure}

\subsection{Comparison with SDO/HMI and Hinode/SOT-SP}

We further evaluated the MVPinn performance by cross-referencing its results with co-temporal magnetograms from the well-established Solar Dynamics Observatory (SDO)/Helioseismic and Magnetic Imager (HMI) and the Hinode Solar Optical Telescope Spectro-Polarimeter (SOT/SP), as shown in Figure \ref{fig:stacked_images}. All three instruments display excellent agreement in capturing the large-scale magnetic structures.

When comparing with the HMI $B_z$ map, MVPinn consistently retrieves sharper fine-scale magnetic structures and exhibits higher contrast, particularly in strong-field regions such as sunspot umbrae. This enhancement is expected given the significantly higher pixel scale of BBSO/NIRIS ($\sim$0.083'') compared to HMI (0.5''). The distribution comparison reveals that MVPinn retrieves a broader range of field strengths, especially near $\pm$2500 G, where HMI tends to under-represent the strongest fields. The two-dimensional histogram shows a strong linear correlation (r = 0.956) between MVPinn and HMI $B_z$ values. MVPinn systematically retrieves slightly stronger magnetic fields than HMI in equivalent regions, likely due to the finer spatial sampling available from ground-based observations and much better spectral resolution of NIRIS than HMI.

The comparison with Hinode/SOT-SP reveals a similarly strong structural agreement. While MVPinn still retrieves slightly stronger fields in high-field regions, the overall magnetic field distribution closely overlaps with that from Hinode. This trend is consistent with expectations and may reflect the deeper formation height and higher spatial resolution of BBSO/NIRIS. The scatter plot shows that most points lie near the diagonal line (r=0.903), indicating good agreement between the two inversion results. Although the distribution correlation is strong, the lower overall correlation compared to the HMI case is likely attributed to Hinode’s long raster scan, which typically spans about an hour and introduces temporal averaging that smooths out dynamic photospheric evolution.

\section{Summary and Discussion}

We presented MVPinn, a physics-informed neural network framework developed for ME inversion of high-resolution spectropolarimetric data from BBSO/NIRIS. By embedding the analytical solution of the polarized radiative transfer equation directly into the loss function, MVPinn preserves physical consistency while leveraging the efficiency of modern machine learning. The model is trained on a single calibrated Stokes profile in approximately two hours and subsequently performs rapid inference at a rate of $\sim$15 seconds per magnetogram for the remainder of the day’s observations. The inferred atmospheric parameters closely follow ME assumptions and remain traceable to the original measurements. Quantitative comparisons with traditional ME inversion across representative regions, along with cross-instrument validation using spaceborne data from SDO/HMI and Hinode/SOT-SP, confirm that MVPinn achieves physically reliable, noise-tolerant inversions with significantly reduced computational cost.

Although MVPinn was adapted specifically for BBSO/NIRIS data, its architecture is generalizable to a wide range of spectropolarimetric inversion problems. The framework establishes a foundation for future extensions beyond the ME approximation, enabling the incorporation of stratified atmospheric models with depth-dependent parameters. Such capabilities would support multi-line inversions spanning various atmospheric heights, i.e., the Fe I 6302 Å line, Ca II 8542 Å line, and He I 10830 Å triplet. 

The practical advantages of MVPinn are significant. Its GPU-accelerated, batch-parallel design enables efficient processing of large-volume, high-cadence datasets, making physics-based inversions feasible for daily observations. The speedup allows full-day high-resolution magnetograms to be inverted in minutes, in contrast to hours or days required by traditional Levenberg–Marquardt–based ME codes. In addition, MVPinn shows strong resilience to observational noise and low-level artifacts (e.g., cross talk), a persistent challenge in classical pixel-by-pixel inversions. Its physics-informed constraints naturally regularize the solution space, suppressing noise-induced fluctuations while retaining genuine polarization signatures, such as weak-field regions where classical approaches often yield unreliable results.

Despite its strengths, limitations remain. The ME approximation assumes a simplified, depth-independent atmosphere, which constrains its applicability in regions with strong vertical gradients or high dynamics. While MVPinn robustly recovers parameters consistent with this approximation, caution should be exercised when interpreting results in stratified or multi-component conditions. Moreover, the model’s generalization capacity depends on the representativeness of the training data. In this study, we assumed relatively stable observing conditions throughout the day. Although daily retraining is a practical solution, transfer learning or partial fine-tuning of a pre-trained model may provide a more efficient adaptation strategy. For future implementation, we aim to evaluate long-term model performance across varying conditions and develop adaptive schemes that minimize retraining while preserving accuracy.

\begin{acknowledgments}
We gratefully acknowledge the use of data from the Goode Solar Telescope (GST) of the Big Bear Solar Observatory (BBSO). BBSO operation is supported by  NSF grants AGS-2309939 and New Jersey Institute of Technology. GST operation is partly supported by the Korea Astronomy and Space Science Institute and the Seoul National University. This work was supported by NSF grant AGS-2309939, and NASA grants,   and 80NSSC24M0174 and 80NSSC24K0258.
\end{acknowledgments}

\bibliography{reference}{}
\bibliographystyle{aasjournalv7}



\end{document}